\pdfoutput=1
\documentclass[twoside]{article}

\usepackage[a4paper]{geometry}
\usepackage[utf8]{inputenc}
\usepackage[T1]{fontenc} 
\usepackage{RR}
\usepackage{hyperref}
\usepackage{amsmath,amssymb,amsfonts}
\usepackage{enumitem}
\usepackage{float} 
\usepackage{multirow}

\newcommand{\conc}{\ensuremath{\,\|\,}}
\newcommand{\Or}{\vee}
\renewcommand{\And}{\wedge}
\newcommand{\card}{\mbox{card\,}}
\newcommand{\I}[1]{\mbox{\it #1}}
\newcommand{\R}[1]{\mbox{\rm #1}}
\newcommand{\ceil}[1]{\lceil #1 \rceil}

\definecolor{grey}{cmyk}{0,0,0,0.7}

\hypersetup{
pdfborder= 0 0 0,
colorlinks=true,
anchorcolor=grey,
citecolor=grey,
filecolor=grey,
linkcolor=grey,
menucolor=grey,
runcolor=grey,
urlcolor=grey
}
\RRdate{December 2021}

\RRauthor{Pierre Bouvier}
\authorhead{P. Bouvier}
\RRtitle{Le jeu de tests VLSAT-3}
\RRetitle{The VLSAT-3 Benchmark Suite}
\titlehead{The VLSAT-3 Benchmark Suite}
\RRresume{VLSAT-3 (acronyme anglais de ``tr\`es grands probl\`emes de
satisfaisabilit\'e bool\'eenne'') est le troisi\`eme volet d'une suite de tests
destin\'ee aux exp\'erimentations scientifiques et aux comp\'etitions
de logiciels pour la r\'esolution de probl\`emes SAT et SMT
(Satisfaisabilit\'e Modulo des Th\'eories).
VLSAT-3 contient 1200~formules logiques (600~satisfaisables
et 600~insatisfaisables) du premier ordre sans quantificateur,
de complexit\'e croissante,
fournies en format SMT-LIB sous une licence Creative Commons permissive.
Plus de 90\% de ces tests ont \'et\'e utilis\'es lors
de la 16\textsuperscript{\`eme} Comp\'etition Internationale
de Satisfaisabilit\'e Modulo des Th\'eories (SMT-COMP~2021).
}
\RRabstract{
This report presents VLSAT-3 (an acronym for ``Very Large Boolean SATisfiability
problems''), the third part of a benchmark suite to be used in scientific
experiments and software competitions addressing SAT and SMT
(Satisfiability Modulo Theories) solving issues.
\mbox{VLSAT-3} contains 1200 (600~satisfiable and 600~unsatisfiable)
quantifier-free first-order logic formulas
of increasing complexity, proposed in SMT-LIB format under a permissive
Creative Commons license. More than 90\% of these benchmarks
have been used during the 16\textsuperscript{th} International Satisfiability Modulo Theories
Competition (SMT-COMP~2021).
}
\RRmotcle{ensemble de donn\'ees, formule SMT, Nested-Unit Petri Net, NUPN, r\'eseau de Petri, r\'esolveur SMT, satisfaisabilit\'e modulo des th\'eories, SMT-LIB, suite de tests}
\RRkeyword{benchmark suite, data set, SMT-LIB, Nested-Unit Petri Net, NUPN, Petri Net, Satisfiability Modulo Theories, SMT formula, SMT solver}
\RRprojet{CONVECS}
\URRhoneAlpes

\begin{document}
\RTNo{0516} 
\makeRT

\section{Benchmark Description}

We previously published two test suites, named VLSAT-1~\cite{Bouvier-Garavel-20}
and VLSAT-2~\cite{Bouvier-Garavel-21-c}, containing SAT formulas.
VLSAT-3\footnote{\url{https://cadp.inria.fr/resources/vlsat/3.html}}
is a collection of 1200~SMT (Satisfiability Modulo Theories) formulas
(i.e., first-order logic formulas),
written in six~different quantifier-free logic fragments.
For each of these logic fragments, 100~satisfiable and 100~unsatisfiable formulas
are provided, which results in 12~families containing 100~benchmarks each.

Each formula is provided as a separate file, encoded in the SMT-LIB~2.6
format~\cite{Barrett-Fontaine-Tinelli-17}.
Each file is then compressed using bzip2 to save disk space and allow
faster downloads. The 1200~formulas require 2.4~gigabytes of disk space
and 132~megabytes when compressed using bzip2.

The VLSAT-3 benchmarks are licensed under the CC-BY Creative Commons
Attribution 4.0 International License\footnote{License terms available from
\url{http://creativecommons.org/licenses/by/4.0}}.

\section{Scientific Context}
\label{CONTEXT}

Interesting SMT formulas can be generated as a by-product of our recent
work \cite{Bouvier-Garavel-PonceDeLeon-20} on the decomposition of Petri nets 
into networks of automata, a problem that has been around since the early 
70s. Concretely, we developed a tool chain that takes as input a
Petri net (which must be ordinary, safe, and hopefully not too large)
and produces as output a network of automata that execute concurrently
and synchronize using shared transitions. Precisely, this network is
expressed as a {\em Nested-Unit Petri Net\/} (NUPN)~\cite{Garavel-19},
i.e., an extension of a Petri net, in which places are grouped into sets
(called {\em units\/}) that denote sequential components. A NUPN provides
a proper structuring of its underlying Petri net, and enables formal 
verification tools to be more efficient in terms of memory and CPU time.
Hence, the NUPN concept has been implemented in many tools
and adopted by software competitions, such as the Model Checking 
Contest\footnote{\url{https://mcc.lip6.fr}}
\cite{Kordon-Garavel-et-al-16,Kordon-Garavel-et-al-18} and the Rigorous
Examination of Reactive Systems challenge\footnote{\url{http://rers-challenge.org}}
\cite{Jasper-Fecke-Steffen-et-al-17,Steffen-Jasper-Meijer-vandePol-17,Jasper-Mues-Murtovi-et-al-19}.
Each NUPN generated by our tool chain is {\em flat}, meaning that its units
are not recursively nested in each other, and {\em unit-safe}, meaning that
each unit has at most one execution token at a time.

Our tool chain works by reformulating concurrency constraints on Petri nets
as logical problems, which can be later solved using third-party software,
such as SAT solvers, SMT solvers, and tools for graph coloring and finding 
maximum cliques, depending on the chosen
strategy~\cite{Bouvier-Garavel-PonceDeLeon-20}.
When a strategy involving SMT solving is selected,
the tool chain produces formulas to be processed by SMT solvers.
The tool chain is solver-agnostic and supports six~standard SMT logic fragments.

\section{Structure of Formulas}
\label{STRUCTURE}

Each of our formulas was produced for a particular Petri net. A formula 
depends on four factors: 
\begin{itemize}
\item the set $P$ of the places of the Petri net;
\item a {\em concurrency relation\/} $\conc$ defined over $P$, such that
$p \conc p'$ iff both places $p$ and $p'$ may simultaneously have 
an execution token;
\item a chosen number $n$ of units; and
\item a chosen SMT logic fragment,
among QF\_BV, QF\_DT, QF\_IDL, QF\_UFBV, QF\_UFDT, and QF\_UFIDL.
\end{itemize} 
A formula checks whether there exists a partition of $P$ into $n$ subsets $P_i$ 
($1 \leq i \leq n$) such that, for each $i$, and for any two places 
$p$ and $p'$ of $P_i$, $p \neq p' \!\!\implies\!\! \neg \, (p \conc p')$.
A model of this formula is thus an allocation of places into $n$ leaf units,
i.e., a valid decomposition of the Petri net.
This can also be seen as an instance of the graph coloring problem, in which 
$n$ colors are to be used for the graph with vertices defined by the places
of $P$ and edges defined by the concurrency relation.
A formula is only satisfiable if the value of $n$ is large enough (namely,
greater than or equal to the chromatic number of the graph), so that at least 
one decomposition exists.

Notice that if a decomposition having $n$ units exists,
it is possible to generate $n! - 1$ another similar
decompositions, just by permuting unit numbers.
Thus, instead of adding constraints in the formulas
to express that each place $p$ belongs to some unit
of $1 \mathrel{{.}\,{.}\,{.}} n$, we break
the symmetry between units by constraining each place $p$ to belongs
some unit of $1 \mathrel{{.}\,{.}\,{.}} \R{min} (p, n)$.

More precisely, each formula is generated as follows, depending on
the chosen logic fragment:

\begin{itemize}
\item The QF\_BV fragment corresponds to {\em quantifier-free bit-vector\/}
logic. It supports fixed-size Boolean vectors, as well as logical, relational,
and arithmetical operators on these vectors.
Our encoding for QF\_BV creates, for each place $p$, a bit
vector $b_p$ of length $n$ such that $b_p[u]$ is true iff place $p$ can
belong to unit $u$.
Then, to prevent concurrent places being assigned in the same unit,
the following constraint is added:
for each pair of places $(p_1, p_2)$ such that
$(p_1 \conc p_2) \And (p_1 < p_2)$,
$b_{p_1} \mathbin{\&} b_{p_2} = 0_{(n)}$, where ``$\mathbin{\&}$''
denotes the bitwise ``and'' operator, and ``$0_{(n)}$'' denotes
the zero vector of size $n$.
Finally, for each place $p$, we add the following symmetry-breaking
constraint to express that $p$ belongs to at least one unit:
$\bigvee_{1 \le u \le \R{min} (p, n)} (b_p[1:u] \neq 0_{(u)})$,
where $[i:j]$ denotes the operator that extracts, from a vector, the
$i$\textsuperscript{th} bit to the $j$\textsuperscript{th}.
Notice that there is no constraint requiring that each place
belongs to one single unit.
Thus, the model returned by the SMT solver may not be a
partition of $P$. Yet, having a partition is necessary to produce
a valid decomposition of the Petri net into a flat NUPN.
To do so, any model that is not a partition is transformed
into a partition using the first-fit-decreasing bin-packing algorithm
described in~\cite{Bouvier-Garavel-PonceDeLeon-20}.

\item The QF\_UFBV fragment corresponds to
{\em quantifier-free uninterpreted-function bit-vector\/} logic.
The {\em uninterpreted function\/} theory enables the declaration of function
symbols, which are given only by their signatures (i.e., the types of their arguments
and results). Our encoding for QF\_UFBV is based on that of QF\_BV but,
instead of the $b_p$ variables,
we define an uninterpreted function \I{u} from the set of bitvectors
of size $\ceil{\log_2(\card(P))}$ to the set of bitvectors
of size $n$, each occurrence of $b_p$ being replaced with
$\I{u} (\lambda(\#p))$ in the constraints, where $\#p$ is a bijection
from places numbers to the interval $1 \mathrel{{.}\,{.}\,{.}} \card(P)$,
and where $\lambda$ is an injection from $1 \mathrel{{.}\,{.}\,{.}} \card(P)$
to the set of bitvectors of size $\ceil{\log_2(\card(P))}$.

\item The QF\_DT fragment corresponds to {\em quantifier-free data-type\/}
logic. It supports the definition of algebraic data types,
such as enumerated types, records, lists, trees, etc.
Our encoding for QF\_DT defines an enumerated type \I{Unit}, which contains one
value per unit. Our encoding also creates, for each place $p$, one variable
$x_p$ of type \I{Unit}.
Then, to prevent concurrent places being assigned to the same unit,
the following constraint is added: for each pair of places
$(p_1, p_2)$ such that $(p_1 \conc p_2) \And (p_1 < p_2)$,
$x_{p_1} \neq x_{p_2}$.
Finally, the symmetry is broken by adding, for each variable $x_p$
whose place number $\#p$ is less than $n$, the constraint
$\bigvee_{1 \le u \le p} (x_p = u)$.

\item The QF\_UFDT fragment corresponds to
{\em quantifier-free uninterpreted-function data-type\/} logic.
Our encoding for QF\_UFDT is based on that of QF\_DT but,
instead of the $x_p$ variables,
defines both an enumerated type \I{Place}, which contains one value per
place, and an uninterpreted function
$\I{u} : \I{Place} \mapsto \I{Unit}$, each occurrence of $x_p$
being replaced with $\I{u} (p)$ in the constraints.

\item The QF\_IDL fragment corresponds to
{\em quantifier-free integer-difference\/} logic.
It supports integer variables and arithmetic constraints on the
difference between two variables.
The {\em integer difference logic\/} theory provides integers, which can also
have constraints of the form $(x - y) \; \I{op} \; c$,
where $x$ is an integer variable, $y$ is either an integer variable
or a constant, \I{op} is a comparison operator,
and $c$ is an integer constant.
Our encoding for QF\_IDL is based on that of QF\_DT but declares
the variables $x_p$ to be of Integer type instead of \I{Unit}.
Since integers are unbounded, each variable $x_p$ whose place number $\#p$
is greater or equal than $n$ must be constrained by adding
$\bigvee_{1 \le u \le n} (x_p = u)$, since $x_p$ is not subject
to a symmetry-break constraint.
Notice that we could replace some disjunctive clauses with difference constraints
(e.g., $x_3 = 1 \Or x_3 = 2 \Or x_3 = 3$ being replaced with
$0 \le x_3 \le 3$); this variant was tried, but found to be slower
during our early experiments using the Z3 solver,
and therefore, not used later.

\item The QF\_UFIDL fragment corresponds to
{\em quantifier-free uninterpreted-function \mbox{integer-difference\/}\/} logic.
Our encoding for QF\_UFIDL is based on that of QF\_IDL,
with the same changes as for evolving from QF\_BV to QF\_UFBV.
\end{itemize}

\section{Selection of Benchmarks}

We applied our approach to a large collection of more than~12,000 Petri nets
from multiple sources, many of which are related to industrial problems,
such as communication protocols, distributed systems, and hardware circuits.
We thus generated a large collection
of more than 51,000~SMT formulas produced by our tool chain.
In this collection, we carefully selected a subset of formulas
matching the requirements of the 16\textsuperscript{th}
International Satisfiability Modulo Theories Competition
(SMT-COMP~2021)\footnote{\url{https://smt-comp.github.io/2021/}}.

For our experiments, we used six state-of-the-art solvers:
Boolector~3.2.0 (compiled with its ``\mbox{-{}-only-cadical}'' option),
Bitwuzla (the version submitted to the SMT-COMP~2020),
CVC4~1.8, MathSAT~5.6.5, Yices~2.6.2, and Z3~4.8.9.
These include all the solvers that participated in \mbox{SMT-COMP~2019}
and/or in SMT-COMP~2020 and have been the fastest or the second fastest
for any logic fragment supported by our toolchain
(i.e., Boolector, Bitwuzla, CVC4, Yices, and~Z3),
as well as the solvers used for constructing
VLSAT-2~\cite{Bouvier-Garavel-21-c} and that can solve both
SAT and SMT formulas (i.e., MathSAT and~Z3).

For each formula $\phi$, let $\R{min}(\phi)$ the smallest execution
time measured among all the six selected solvers, using a Linux server
equipped with a Xeon~E5-2630~v3 and 128~GB~RAM. We selected
the 46,000+ formulas $\phi$ such that $\R{min}(\phi)$ ranges
between 10~seconds and 1~hour.
For each of the six supported logic fragments,
we created one family of satisfiable formulas
and one family of unsatisfiable formulas.
We partitioned each family in (at most) 60~classes,
by gathering in the same class all formulas $\phi$ having
the same $\R{min}(\phi)$ rounded to the nearest minute.
Finally, in each class, we selected a few formulas
having a low complexity
(measured in terms of SMT-LIB file size),
and took the same number (plus or minus one) of formulas
in each class, until reaching 100~formulas per family.

\section{Resulting Benchmarks}

The 12 families of the VLSAT-3 benchmark suite are listed in
Table~\ref{TABLE-FAMILIES}, the columns of which have the following meaning:

\begin{itemize}
\item Column~1: Letter identifying the corresponding family of formulas.
\item Column~2: Satisfiability (SAT or UNSAT);
all formulas in a given family have the same satisfiability.
\item Column~3: One out of six quantifier-free first-order logic fragment;
all formulas in a given family are in the same logic fragment.
\item Column~4: Percentage of the formulas in this family
that have been selected by the organizers of SMT-COMP~2021.
\item Column~5: Percentage of the formulas in this family
among all the benchmarks of the corresponding logic
in the Single Query Track (SQT) of SMT-COMP~2021.
\item Column~6: Percentage of the formulas in this family
among all the benchmarks of the corresponding logic
in the Model Validation Track (MVT) of SMT-COMP~2021.
\item Column~7: Percentage of the formulas in this family
among all the benchmarks of the corresponding logic
in the Unsat Core Track (UCT) of SMT-COMP~2021.
\end{itemize}

\begin{table}[h]
\centering
\begin{tabular}{|c|c|l|r|r|r|r|}
\hline
family & satisfiability & logic & selection & SQT share & MVT share & UCT share \\
\hline
a & \multirow{6}{*}{UNSAT}  & QF\_BV    & 76~\%  & 0.86~\%   &          & 2.65~\%  \\
b &                         & QF\_DT    & 100~\% & 49.02~\%  &          & 100~\%   \\
c &                         & QF\_IDL   & 100~\% & 8.79~\%   &          & 100~\%   \\
d &                         & QF\_UFBV  & 88~\%  & 29.33~\%  &          & 30.67~\% \\
e &                         & QF\_UFDT  & 100~\% & 49.26~\%  &          & 100~\%   \\
f &                         & QF\_UFIDL & 100~\% & 33.33~\%  &          & 88.50~\% \\
\hline
g & \multirow{6}{*}{SAT}    & QF\_BV    & 76~\%  & 0.86~\%   & 1.05~\%  &         \\
h &                         & QF\_DT    & 100~\% & 49.02~\%  &          &         \\
i &                         & QF\_IDL   & 70~\%  & 6.15~\%   & 10.13~\% &         \\
j &                         & QF\_UFBV  & 81~\%  & 27.0~\%   & 21.6~\%  &         \\
k &                         & QF\_UFDT  & 100~\% & 49.26~\%  &          &         \\
l &                         & QF\_UFIDL & 100~\% & 33.33~\%  & 48.54~\% &         \\
\hline
\end{tabular}
\caption{\label{TABLE-FAMILIES} List of the 12 families of VLSAT-3 formulas}
\end{table}

In a nutshell, 90.9~\% of the VLSAT-3 benchmarks have been used
by the organizers of \mbox{SMT-COMP~2021}. Additionally,
7.8~\%~of the benchmarks of the Single Query Track,
2.9~\%~of the benchmarks of the Model Validation Track,
and 14.0~\%~of the benchmarks of the Unsat Core Track are VLSAT-3 benchmarks.

We now provide 12~tables that describe each family in detail.
The first one, Table~\ref{TABLE-A}, corresponds to family~``a'',
and the last one, Table~\ref{TABLE-L}, corresponds to family~``l''.

Each row of a table corresponds to one benchmark.
The columns of these tables contain the following data:
\begin{itemize}
\item Column name: Benchmark name, of the form ``vlsat3\_Fnn.smt2.bz2'',
where F is the family letter, and nn is a two-digit number ranging from 00 to 99.
\item Column \#variables (for QF\_BV, QF\_DT, and QF\_IDL):
Number of variables declared in the formula.
\item Column card (for QF\_BV, QF\_DT, and QF\_IDL):
Cardinality of the type of all variables declared in the formula
(or $\infty$ if the type has an infinite set of values).
\item Column card\textsubscript{in}:
(for QF\_UFBV, QF\_UFDT, and QF\_UFIDL):
Cardinality of the domain of the uninterpreted function present in the formula
(or $\infty$ if the domain is infinite).
\item Column card\textsubscript{out}:
(for QF\_UFBV, QF\_UFDT, and QF\_UFIDL):
Cardinality of the codomain of the uninterpreted function present in the formula
(or $\infty$ if the codomain is infinite).
\item Column \#asserts: Number of assertions in the formula.
\item Column \#ops: Number of operators in the formula.
\end{itemize}

\subsubsection*{Acknowledgements}
The experiments presented in this paper were carried out using the 
{\sc Grid'5000}\footnote{\url{https://www.grid5000.fr}} testbed, supported 
by a scientific interest group hosted by {\sc Inria} and including {\sc Cnrs},
{\sc Renater} and several universities as well as other organizations. 

\clearpage

\begin{table}[H]
\centering
\setlength{\tabcolsep}{5pt}
\small
\caption{\label{TABLE-A} Family ``a'': List of unsatisfiable VLSAT-3 formulas written in QF\_BV logic}
\vspace{5pt}
\begin{tabular}{|c|r|r|r|r|} \hline
name & \#variables & card & \#asserts & \#ops \\ \hline \hline
a00 & 56  & $2^{11}$  & 1463    & 7283       \\
a01 & 80  & $2^{11}$  & 3050    & 15,170     \\
a02 & 80  & $2^{17}$  & 3050    & 15,218     \\
a03 & 56  & $2^{12}$  & 1463    & 7291       \\
a04 & 120 & $2^{15}$  & 6975    & 34,747     \\
a05 & 140 & $2^{17}$  & 8750    & 43,598     \\
a06 & 80  & $2^{18}$  & 3050    & 15,226     \\
\hline
\end{tabular}
\hfill
\begin{tabular}{|c|r|r|r|r|} \hline
name & \#variables & card & \#asserts & \#ops \\ \hline \hline
a50 & 140 & $2^{19}$  & 8750    & 43,614     \\
a51 & 114 & $2^{11}$  & 2465    & 12,177     \\
a52 & 90  & $2^{18}$  & 2808    & 13,996     \\
a53 & 118 & $2^{18}$  & 6333    & 31,565     \\
a54 & 176 & $2^{11}$  & 12,053  & 59,993     \\
a55 & 47  & $2^{16}$  & 745     & 3751       \\
a56 & 57  & $2^{13}$  & 1306    & 6512       \\
\hline
\end{tabular}
\end{table}

\begin{table}[H]
\centering
\setlength{\tabcolsep}{5pt}
\small
\caption{\label{TABLE-B} Family ``b'': List of unsatisfiable VLSAT-3 formulas written in QF\_DT logic}
\vspace{5pt}
\begin{tabular}{|c|r|r|r|r|} \hline
name & \#variables & card & \#asserts & \#ops \\ \hline \hline
b00      & 56          & 16   & 1422      & 4504        \\
b01      & 164         & 11   & 13,018    & 39,162      \\
b02      & 56          & 12   & 1418      & 4384        \\
b03      & 120         & 16   & 6870      & 20,848      \\
b04      & 118         & 14   & 4890      & 14,850      \\
b05      & 140         & 18   & 8627      & 26,185      \\
b06      & 155         & 13   & 9185      & 27,709      \\
\hline
\end{tabular}
\hfill
\begin{tabular}{|c|r|r|r|r|} \hline
name & \#variables & card & \#asserts & \#ops \\ \hline \hline
b50      & 74          & 13   & 1844      & 5686        \\
b51      & 141         & 11   & 5923      & 17,877      \\
b52      & 187         & 11   & 8484      & 25,560      \\
b53      & 163         & 14   & 9451      & 28,533      \\
b54      & 263         & 11   & 18,641    & 56,031      \\
b55      & 108         & 11   & 2849      & 8655        \\
b56      & 80          & 19   & 2988      & 9304        \\
\hline
\end{tabular}
\end{table}

\begin{table}[H]
\centering
\setlength{\tabcolsep}{5pt}
\small
\caption{\label{TABLE-C} Family ``c'': List of unsatisfiable VLSAT-3 formulas written in QF\_IDL logic}
\vspace{5pt}
\begin{tabular}{|c|r|r|r|r|} \hline
name & \#variables & card & \#asserts & \#ops \\ \hline \hline
c00      & 56          & $\infty$    & 1463      & 5687        \\
c01      & 80          & $\infty$    & 3050      & 10,936      \\
c02      & 215         & $\infty$    & 19,420    & 61,226      \\
c03      & 216         & $\infty$    & 23,436    & 117,176     \\
c04      & 56          & $\infty$    & 1463      & 5773        \\
c05      & 80          & $\infty$    & 3050      & 11,072      \\
c06      & 80          & $\infty$    & 3050      & 11,206      \\
\hline
\end{tabular}
\hfill
\begin{tabular}{|c|r|r|r|r|} \hline
name & \#variables & card & \#asserts & \#ops \\ \hline \hline
c50      & 272         & $\infty$    & 18,443    & 62,243      \\
c51      & 297         & $\infty$    & 22,683    & 75,613      \\
c52      & 80          & $\infty$    & 3050      & 11,596      \\
c53      & 120         & $\infty$    & 6975      & 24,523      \\
c54      & 136         & $\infty$    & 8792      & 33,016      \\
c55      & 247         & $\infty$    & 15,746    & 53,502      \\
c56      & 55          & $\infty$    & 1026      & 4750        \\
\hline
\end{tabular}
\end{table}

\begin{table}[H]
\centering
\setlength{\tabcolsep}{5pt}
\small
\caption{\label{TABLE-D} Family ``d'': List of unsatisfiable VLSAT-3 formulas written in QF\_UFBV logic}
\vspace{5pt}
\begin{tabular}{|c|r|r|r|r|} \hline
name & card\textsubscript{in} & card\textsubscript{out} & \#asserts & \#ops \\ \hline \hline
d00      & $2^{7}$    & $2^{17}$    & 3050      & 21,142      \\
d01      & $2^{8}$    & $2^{14}$    & 8576      & 59,647      \\
d02      & $2^{7}$    & $2^{16}$    & 6975      & 48,495      \\
d03      & $2^{8}$    & $2^{15}$    & 8576      & 59,649      \\
d04      & $2^{7}$    & $2^{12}$    & 1819      & 12,503      \\
d05      & $2^{7}$    & $2^{19}$    & 3050      & 21,146      \\
d06      & $2^{6}$    & $2^{13}$    & 1407      & 9693        \\
\hline
\end{tabular}
\hfill
\begin{tabular}{|c|r|r|r|r|} \hline
name & card\textsubscript{in} & card\textsubscript{out} & \#asserts & \#ops \\ \hline \hline
d50      & $2^{7}$    & $2^{13}$    & 2508      & 17,319      \\
d51      & $2^{8}$    & $2^{13}$    & 7304      & 50,723      \\
d52      & $2^{6}$    & $2^{13}$    & 591       & 4032        \\
d53      & $2^{7}$    & $2^{12}$    & 3741      & 25,891      \\
d54      & $2^{6}$    & $2^{21}$    & 1113      & 7687        \\
d55      & $2^{6}$    & $2^{24}$    & 1506      & 10,423      \\
d56      & $2^{7}$    & $2^{13}$    & 6363      & 44,223      \\
\hline
\end{tabular}
\end{table}

\begin{table}[H]
\centering
\setlength{\tabcolsep}{5pt}
\small
\caption{\label{TABLE-E} Family ``e'': List of unsatisfiable VLSAT-3 formulas written in QF\_UFDT logic}
\vspace{5pt}
\begin{tabular}{|c|r|r|r|r|} \hline
name & card\textsubscript{in} & card\textsubscript{out} & \#asserts & \#ops \\ \hline \hline
e00      & 56     & 16      & 1422      & 7438        \\
e01      & 80     & 15      & 2984      & 15,205      \\
e02      & 80     & 16      & 2985      & 15,253      \\
e03      & 56     & 12      & 1418      & 7264        \\
e04      & 164    & 11      & 13,018    & 65,233      \\
e05      & 196    & 11      & 18,721    & 93,748      \\
e06      & 70     & 14      & 1576      & 8125        \\
\hline
\end{tabular}
\hfill
\begin{tabular}{|c|r|r|r|r|} \hline
name & card\textsubscript{in} & card\textsubscript{out} & \#asserts & \#ops \\ \hline \hline
e50      & 74     & 13      & 1844      & 9428        \\
e51      & 141    & 11      & 5923      & 29,758      \\
e52      & 169    & 11      & 7133      & 35,808      \\
e53      & 201    & 17      & 12,460    & 62,674      \\
e54      & 41     & 14      & 501       & 2750        \\
e55      & 87     & 14      & 2598      & 13,235      \\
e56      & 174    & 21      & 9097      & 46,073      \\
\hline
\end{tabular}
\end{table}

\begin{table}[H]
\centering
\setlength{\tabcolsep}{5pt}
\small
\caption{\label{TABLE-F} Family ``f'': List of unsatisfiable VLSAT-3 formulas written in QF\_UFIDL logic}
\vspace{5pt}
\begin{tabular}{|c|r|r|r|r|} \hline
name & card\textsubscript{in} & card\textsubscript{out} & \#asserts & \#ops \\ \hline \hline
f00      & $\infty$      & $\infty$       & 1463      & 9406        \\
f01      & $\infty$      & $\infty$       & 13,172    & 70,315      \\
f02      & $\infty$      & $\infty$       & 18,871    & 112,070     \\
f03      & $\infty$      & $\infty$       & 3050      & 18,949      \\
f04      & $\infty$      & $\infty$       & 6975      & 39,718      \\
f05      & $\infty$      & $\infty$       & 8750      & 50,200      \\
f06      & $\infty$      & $\infty$       & 1463      & 9151        \\
\hline
\end{tabular}
\hfill
\begin{tabular}{|c|r|r|r|r|} \hline
name & card\textsubscript{in} & card\textsubscript{out} & \#asserts & \#ops \\ \hline \hline
f50      & $\infty$      & $\infty$       & 7816      & 43,871      \\
f51      & $\infty$      & $\infty$       & 1140      & 8560        \\
f52      & $\infty$      & $\infty$       & 1245      & 8468        \\
f53      & $\infty$      & $\infty$       & 1642      & 10,313      \\
f54      & $\infty$      & $\infty$       & 320       & 2400        \\
f55      & $\infty$      & $\infty$       & 1778      & 11,856      \\
f56      & $\infty$      & $\infty$       & 21,296    & 135,145     \\
\hline
\end{tabular}
\end{table}

\begin{table}[H]
\centering
\setlength{\tabcolsep}{5pt}
\small
\caption{\label{TABLE-G} Family ``g'': List of satisfiable VLSAT-3 formulas written in QF\_BV logic}
\vspace{5pt}
\begin{tabular}{|c|r|r|r|r|} \hline
name & \#variables & card & \#asserts & \#ops \\ \hline \hline
g00      & 127         & $2^{46}$  & 7987      & 40,041      \\
g01      & 238         & $2^{27}$  & 9157      & 45,517      \\
g02      & 160         & $2^{62}$  & 12,490    & 62,618      \\
g03      & 160         & $2^{63}$  & 12,490    & 62,626      \\
g04      & 224         & $2^{89}$  & 15,832    & 79,416      \\
g05      & 192         & $2^{93}$  & 18,397    & 92,337      \\
g06      & 200         & $2^{80}$  & 19,735    & 98,907      \\
\hline
\end{tabular}
\hfill
\begin{tabular}{|c|r|r|r|r|} \hline
name & \#variables & card & \#asserts & \#ops \\ \hline \hline
g50      & 273         & $2^{52}$  & 29,938    & 149,552     \\
g51      & 299         & $2^{53}$  & 38,293    & 191,283     \\
g52      & 353         & $2^{52}$  & 45,498    & 227,192     \\
g53      & 459         & $2^{47}$  & 49,154    & 245,220     \\
g54      & 325         & $2^{66}$  & 52,331    & 261,525     \\
g55      & 120         & $2^{60}$  & 7029      & 35,377      \\
g56      & 157         & $2^{56}$  & 12,212    & 61,186      \\
\hline
\end{tabular}
\end{table}

\begin{table}[H]
\centering
\setlength{\tabcolsep}{5pt}
\small
\caption{\label{TABLE-H} Family ``h'': List of satisfiable VLSAT-3 formulas written in QF\_DT logic}
\vspace{5pt}
\begin{tabular}{|c|r|r|r|r|} \hline
name & \#variables & card & \#asserts & \#ops \\ \hline \hline
h00      & 120         & 84   & 6938      & 27,784      \\
h01      & 138         & 119  & 9497      & 42,531      \\
h02      & 138         & 118  & 9496      & 42,292      \\
h03      & 138         & 113  & 9491      & 41,127      \\
h04      & 215         & 124  & 19,328    & 73,234      \\
h05      & 200         & 129  & 19,553    & 75,169      \\
h06      & 200         & 130  & 19,554    & 75,430      \\
\hline
\end{tabular}
\hfill
\begin{tabular}{|c|r|r|r|r|} \hline
name & \#variables & card & \#asserts & \#ops \\ \hline \hline
h50      & 279         & 29   & 17,595    & 53,595      \\
h51      & 282         & 31   & 25,260    & 76,708      \\
h52      & 244         & 32   & 15,921    & 48,753      \\
h53      & 233         & 32   & 23,702    & 72,096      \\
h54      & 172         & 44   & 14,208    & 44,514      \\
h55      & 408         & 407  & 67,406    & 367,458     \\
h56      & 248         & 101  & 19,541    & 68,721      \\
\hline
\end{tabular}
\end{table}

\begin{table}[H]
\centering
\setlength{\tabcolsep}{5pt}
\small
\caption{\label{TABLE-I} Family ``i'': List of satisfiable VLSAT-3 formulas written in QF\_IDL logic}
\vspace{5pt}
\begin{tabular}{|c|r|r|r|r|} \hline
name & \#variables & card & \#asserts & \#ops \\ \hline \hline
i00      & 215         & $\infty$    & 19,420    & 62,056      \\
i01      & 236         & $\infty$    & 19,484    & 61,252      \\
i02      & 228         & $\infty$    & 19,386    & 86,806      \\
i03      & 200         & $\infty$    & 19,625    & 88,973      \\
i04      & 160         & $\infty$    & 12,490    & 53,526      \\
i05      & 273         & $\infty$    & 29,938    & 115,552     \\
i06      & 150         & $\infty$    & 11,175    & 46,073      \\
\hline
\end{tabular}
\hfill
\begin{tabular}{|c|r|r|r|r|} \hline
name & \#variables & card & \#asserts & \#ops \\ \hline \hline
i50      & 704         & $\infty$    & 245,856   & 1,176,046   \\
i51      & 802         & $\infty$    & 309,062   & 1,427,170   \\
i52      & 800         & $\infty$    & 314,447   & 1,512,047   \\
i53      & 252         & $\infty$    & 31,647    & 139,789     \\
i54      & 511         & $\infty$    & 130,476   & 653,056     \\
i55      & 834         & $\infty$    & 218,412   & 683,318     \\
i56      & 813         & $\infty$    & 300,190   & 1,493,444   \\
\hline
\end{tabular}
\end{table}

\begin{table}[H]
\centering
\setlength{\tabcolsep}{5pt}
\small
\caption{\label{TABLE-J} Family ``j'': List of satisfiable VLSAT-3 formulas written in QF\_UFBV logic}
\vspace{5pt}
\begin{tabular}{|c|r|r|r|r|} \hline
name & card\textsubscript{in} & card\textsubscript{out} & \#asserts & \#ops \\ \hline \hline
j00      & $2^{6}$     & $2^{30}$     & 1000      & 6908        \\
j01      & $2^{6}$     & $2^{25}$     & 1250      & 8648        \\
j02      & $2^{6}$     & $2^{30}$     & 1280      & 8853        \\
j03      & $2^{6}$     & $2^{30}$     & 1665      & 11,533      \\
j04      & $2^{6}$     & $2^{30}$     & 1719      & 11,911      \\
j05      & $2^{7}$     & $2^{40}$     & 4770      & 33,168      \\
j06      & $2^{11}$    & $2^{98}$     & 6048      & 42,200      \\
\hline
\end{tabular}
\hfill
\begin{tabular}{|c|r|r|r|r|} \hline
name & card\textsubscript{in} & card\textsubscript{out} & \#asserts & \#ops \\ \hline \hline
j50      & $2^{10}$    & $2^{77}$     & 62,937    & 439,427     \\
j51      & $2^{10}$    & $2^{78}$     & 62,937    & 439,429     \\
j52      & $2^{9}$     & $2^{14}$     & 73,016    & 509,698     \\
j53      & $2^{9}$     & $2^{56}$     & 76,114    & 531,681     \\
j54      & $2^{9}$     & $2^{56}$     & 76,216    & 532,395     \\
j55      & $2^{10}$    & $2^{61}$     & 77,529    & 541,146     \\
j56      & $2^{10}$    & $2^{63}$     & 82,118    & 573,207     \\
\hline
\end{tabular}
\end{table}

\begin{table}[H]
\centering
\setlength{\tabcolsep}{5pt}
\small
\caption{\label{TABLE-K} Family ``k'': List of satisfiable VLSAT-3 formulas written in QF\_UFDT logic}
\vspace{5pt}
\begin{tabular}{|c|r|r|r|r|} \hline
name & card\textsubscript{in} & card\textsubscript{out} & \#asserts & \#ops \\ \hline \hline
k00 & 138    & 137     & 9515      & 75,249      \\
k01 & 138    & 118     & 9496      & 67,953      \\
k02 & 138    & 115     & 9493      & 66,900      \\
k03 & 164    & 125     & 13,132    & 88,660      \\
k04 & 200    & 126     & 19,550    & 121,123     \\
k05 & 200    & 127     & 19,551    & 121,504     \\
k06 & 200    & 136     & 19,560    & 125,068     \\
\hline
\end{tabular}
\hfill
\begin{tabular}{|c|r|r|r|r|} \hline
name & card\textsubscript{in} & card\textsubscript{out} & \#asserts & \#ops \\ \hline \hline
k50 & 233    & 32      & 23,648    & 119,664     \\
k51 & 185    & 38      & 16,693    & 85,498      \\
k52 & 258    & 41      & 21,181    & 108,283     \\
k53 & 400    & 399     & 79,897    & 636,890     \\
k54 & 621    & 95      & 149,204   & 759,225     \\
k55 & 621    & 112     & 149,221   & 764,529     \\
k56 & 201    & 79      & 12,522    & 71,695      \\
\hline
\end{tabular}
\end{table}

\begin{table}[H]
\centering
\setlength{\tabcolsep}{5pt}
\small
\caption{\label{TABLE-L} Family ``l'': List of satisfiable VLSAT-3 formulas written in QF\_UFIDL logic}
\vspace{5pt}
\begin{tabular}{|c|r|r|r|r|} \hline
name & card\textsubscript{in} & card\textsubscript{out} & \#asserts & \#ops \\ \hline \hline
l00      & $\infty$      & $\infty$       & 19,625    & 144,910     \\
l01      & $\infty$      & $\infty$       & 19,625    & 145,189     \\
l02      & $\infty$      & $\infty$       & 19,625    & 147,565     \\
l03      & $\infty$      & $\infty$       & 19,625    & 153,889     \\
l04      & $\infty$      & $\infty$       & 19,625    & 156,805     \\
l05      & $\infty$      & $\infty$       & 19,625    & 157,978     \\
l06      & $\infty$      & $\infty$       & 13,890    & 158,430     \\
\hline
\end{tabular}
\hfill
\begin{tabular}{|c|r|r|r|r|} \hline
name & card\textsubscript{in} & card\textsubscript{out} & \#asserts & \#ops \\ \hline \hline
l50      & $\infty$      & $\infty$       & 35,153    & 217,390     \\
l51      & $\infty$      & $\infty$       & 45,762    & 378,581     \\
l52      & $\infty$      & $\infty$       & 21,733    & 138,493     \\
l53      & $\infty$      & $\infty$       & 164,523   & 1,555,162   \\
l54      & $\infty$      & $\infty$       & 38,293    & 235,745     \\
l55      & $\infty$      & $\infty$       & 77,738    & 632,091     \\
l56      & $\infty$      & $\infty$       & 46,406    & 379,702     \\
\hline
\end{tabular}
\end{table}

\begin{small}

\end{small}

\end{document}